\documentclass{amsart}

\usepackage{amsmath}
\usepackage{amsfonts}
\usepackage{amssymb}
\usepackage{amsthm}
\usepackage{array}
\usepackage{bbm}
\usepackage{chngpage}
\usepackage{comment}
\usepackage{float}
\usepackage[OT2,T1]{fontenc}
\usepackage{graphicx,caption}
\usepackage[export]{adjustbox}
\usepackage{longtable}
\usepackage{mathrsfs}
\usepackage{mathtools}
\usepackage{wrapfig}
\usepackage{cutwin}
\usepackage{shapepar}
\usepackage{tikz}
\usepackage[lowtilde]{url}
\usepackage{subcaption}
\usetikzlibrary{matrix,arrows}
\usepackage{tabularx}
\usepackage{multirow}
\usepackage[colorlinks, citecolor = blue]{hyperref}
\usepackage[capitalize, nameinlink]{cleveref}
\usepackage{enumitem}
\usepackage{diagbox}
\usepackage{etoolbox}
\usepackage{algpseudocode}

\DeclareMathAlphabet{\mathsfit}{T1}{\sfdefault}{\mddefault}{\sldefault}
\SetMathAlphabet{\mathsfit}{bold}{T1}{\sfdefault}{\bfdefault}{\sldefault}

\usepackage[margin=1.25in]{geometry}

\newcommand{\R}{\mathbb{R}}
\newcommand{\Z}{\mathbb{Z}}

\newcommand{\cN}{\mathcal{N}}
\newcommand{\cO}{\mathcal{O}}
\newcommand{\cP}{\mathcal{P}}

\newcommand{\cS}{\mathcal{S}}

\newcommand{\rnum}[1]{\MakeUppercase{\romannumeral #1}}

\newcommand{\rtwo}{\rnum{1}\hspace{-.7pt}\rnum{1}}

\DeclareSymbolFont{cyrletters}{OT2}{wncyr}{m}{n}
\DeclareMathSymbol{\sha}{\mathalpha}{cyrletters}{"58}

\newcommand{\eps}{\varepsilon}

\newcommand{\ds}{\displaystyle}
\newlength{\strutheight}
\newcommand{\half}{\frac{1}{2}}
\newcommand{\thalf}{\tfrac{1}{2}}

\newtheorem{theorem}{Theorem}[section]

\newtheorem{observation}[theorem]{Observation}
\newtheorem{model}[theorem]{Model}

\theoremstyle{definition}
\newtheorem{remark}[theorem]{Remark}
\newtheorem{algorithm}[theorem]{Algorithm}
\newtheorem{figurecap}[theorem]{Figure}
\newtheorem{definition}[theorem]{Definition}
\newtheorem{example}[theorem]{Example}
\AtBeginEnvironment{example}{%
  \pushQED{\qed}%
}
\AtEndEnvironment{example}{\popQED\endexample}

\author{Alex Cowan}
\address{Department of Mathematics, Harvard University, Cambridge, MA 02138 USA}
\email{cowan@math.harvard.edu}
\thanks{The author was supported by the Simons Foundation Collaboration Grant 550031.}

\title{Paired comparisons for games of chance}
\date{\today}

\begin{document}

\maketitle

\begin{abstract}
  We present a Bayesian rating system based on the method of paired comparisons. Our system is a flexible generalization of the well-known Glicko, and in particular can better accommodate games with significant elements of luck. Our system is currently in use in the online game \textit{Duelyst \rtwo}, and in that setting outperforms Glicko2. 
\end{abstract}

\tableofcontents

\section{Introduction}\label{section:intro}

\textit{The method of paired comparisons} \cite{david} is a framework for ranking many items by comparing them two at a time. Often the outcome of these comparisons is non-deterministic, only a small fraction of all possible pairs will be compared, and some pairs may be compared multiple times. In this paper we discuss the method of paired comparisons in the context of ranking the players of a symmetric competitive two-player game according to their \textit{strength}, i.e.\ ability to win matches. We present a rating system which, based only on the outcomes of previously played matches, estimates how likely any player is to defeat any other player.

There are several other systems designed for this purpose, such as Elo \cite{elo}, Glicko2 \cite{glicko2}, and TrueSkill \cite{trueskill, trueskill2}. Our system is fundamentally a generalization of the well-known system Glicko \cite{glicko} (but not Glicko2; see \cref{remark:glicko2}). With this generalization, we can adapt our system to specific situations in a way that avoids certain assumptions and certain approximations which might not be appropriate in those settings. In particular, Glicko and many other systems use the \textit{Bradley--Terry model} \cite{zermelo, BT} for estimating the winning chances of players whose strength is known exactly. This model tends to overestimate the winning chances of players much stronger than their opponents, and also reflects reality poorly when used in games with elements of luck; c.f.\ \cref{example:randochess} and \cref{subsection:practical_lambda}. Our system is in use in the online collectible card game Duelyst \rtwo, and there our system outperforms Glicko2, which was the rating system the game used previously. We discuss this in \cref{section:performance}.

The core part of our system is made up of three models: \cref{model:match} is used to predict the outcome of matches, \cref{model:update} is used to update the system's beliefs about players based on match outcomes, and \cref{model:evolution} is used to account for changes in player strength between matches due to external factors. The system functions by choosing some prior for a player's strength to assign to unknown players, and then using \cref{model:update} and \cref{model:evolution} after each match.

Models \ref{model:match}, \ref{model:update}, and \ref{model:evolution} are presented in substantial generality, and to use our system one must choose values for three parameters: $\Lambda$ in \cref{model:match}, a prior $\nu_0$ for unknown players to use in \cref{model:update}, and $\kappa$ in \cref{model:evolution}. In \cref{section:practical}, we present the parameter choices we made for Duelyst \rtwo, and give a very succinct summary of the resulting system in \cref{subsection:practical_summary}. We think similar choices will be reasonable in many other situations.

An essential component of Glicko2's popularity is that it is computationally feasible to use it in large scale applications, such as on the major chess website Lichess \cite{lichess:rating}. We present two sets of algorithms in \cref{subsection:algorithms_fft} and \cref{subsection:algorithms_laplace} for implementing our system. These algorithms are efficient enough to be practical for similar applications; Duelyst \rtwo's implementation can process roughly $170$ matches per second per vCPU. We also highlight the algorithms in \cref{subsection:algorithms_naive}, which might help one in understanding the system, and \cref{algorithm:S} and \cref{algorithm:T} for taking convolutions of discrete distributions with Laplace distribution PDFs and CDFs which might be of independent interest. Implementation of all of these algorithms are available on the author's GitHub \cite{github:blatmmr}.

\subsection*{Acknowledgments}
We thank Michael Snarski for many very helpful conversations, Oleg Maslennikov for integrating all aspects of our system into Duelyst \rtwo, and the Duelyst \rtwo\hspace{0pt} community for their enthusiastic testing and thorough feedback.

\section{Model}\label{section:model}
Our system is made up of three models. \cref{model:match} is used to predict the outcome of matches. \cref{model:update} is used after every match to update the system's beliefs about the strengths of players. \cref{model:evolution} is used after each of these updates, to account for changes in player strength between matches due to external factors.

In this section, we present these models in general. To use our system in practice, one then has to choose values for the following parameters:
\begin{itemize}
  \item In \cref{model:match}, $\Lambda$ 
  \item In \cref{model:update}, a prior $\nu_0$ for unknown players
  \item In \cref{model:evolution}, $\kappa$
\end{itemize}
We give a variety of examples of parameter choices for each individual model to highlight relationships to existing systems:
\begin{itemize}
  \item The Bradley--Terry model \cite{zermelo, BT} in \cref{example:BT} and \cref{example:linear_models}
  \item TrueSkill \cite{trueskill, trueskill2} in \cref{example:trueskill}
  \item FIDE \cite{fide} in \cref{example:fide}
  \item Glicko \cite{glicko} in \cref{example:glicko_update} and \cref{example:glicko_kernel}
\end{itemize}
In \cref{section:practical}, we give recommendations for parameter choices which we expect to be suitable for most applications, and which allow for a more concrete formulation of our system.

\subsection{Match outcomes}\label{section:match_outcomes}
The outcome of a match between two players $A$ and $B$ depends on both how well $A$ and $B$ perform in that match as well as the nature of the game that they are playing. In this section we present \cref{model:match} for predicting match outcomes in a way that takes into consideration both of these factors.
\begin{model}\label{model:match}~
  \begin{itemize}
  \item Let $\cS$ be a complete separable metric space.
  \item Let $\cP(\cS)$ be the set of Borel probability measures on $\cS$.
  \item Let $\Lambda: \cS^2 \to [0,1] \subset \R$ be measurable and such that $\Lambda(x,y) = 1 - \Lambda(y,x)$ for all $x,y \in \cS$.
  \item For any $\mu_A$ and $\mu_B$ in $\cP(\cS)$, define
    \begin{align*}
      L(\mu_A,\mu_B) \coloneqq \int_{\cS^2} \Lambda(x,y) \,\mu_A(dx) \,\mu_B(dy).
    \end{align*}
  \end{itemize}
  
  Players $A$ and $B$ are modeled by the probability measures $\mu_A$ and $\mu_B$ respectively, and the average score of $A$ playing against $B$ is modeled by $L(\mu_A,\mu_B)$.
\end{model}

We call $\Lambda$ a \textit{luck function}. In \cref{model:match}, games are determined by the choice of $\Lambda$, and players by their associated probability measures. The luck function $\Lambda$ reflects the nature of the game being played, in particular how much luck influences the outcomes of matches, and should be chosen on a case-by-case basis. The probability measures associated to players reflect how consistently that player performs from match to match.

When two fixed players $A$ and $B$ play multiple matches against one another and multiple outcomes are observed, in many situations it is possible to explain this variation in outcome equally well as a consequence of inconsistent performances by the players (reflected in $\mu_A$ and $\mu_B$), as a consequence of an element of luck inherent in the game (reflected in $\Lambda$), or by various combinations of these two factors. \cref{example:BT} and \cref{example:linear_models} show how the Bradley--Terry model \cite{zermelo, BT} can arise from either the perspective of the game involving luck or players performing inconsistently.

\begin{definition}\label{def:heaviside}
  We take \textit{Heaviside function} to be the function $H:\R \to \R$ defined as
  \begin{align*}
    H(x) =
    \begin{cases}
      0, & x < 0 \\
      \half, & x = 0 \\
      1, & x > 0.
    \end{cases}
  \end{align*}
\end{definition}

\begin{definition}\label{def:dirac_measure}
  For any $x \in \cS$, the \textit{Dirac measure} $\delta_x \in \cP(\cS)$ is defined by
  \begin{align*}
    \delta_x(U) =
    \begin{cases}
      0, & x \not\in U \\
      1, & x \in U
    \end{cases}
  \end{align*}
  for all Borel-measurable sets $U \subseteq \cS$.
\end{definition}

\begin{example}\label{example:basic_match}
  Take $\cS = \R$. If
  \begin{align*}
    \Lambda(x,y) = H(x-y), 
  \end{align*}
  where $H$ is the Heaviside function as in \cref{def:heaviside}, then \cref{model:match} can be interpreted as a game in which the player who performs better in any given match wins that match. Conversely, if
  \begin{align*}
    \Lambda(x,y) = \half \quad\text{for all $x, y \in \cS$},
  \end{align*}
  then this can be interpreted as a game in which the outcome of every match is determined uniformly at random.
  
  If $\mu_A$ is a Dirac measure as in \cref{def:dirac_measure}, i.e.\ $\mu_A(\{x\}) = 1$ for some $x \in \cS$, this can be interpreted as the player $A$ being perfectly consistent, performing with exactly the same strength every match. Conversely, if $\mu_A$ is a probability measure with high variance, then this can be interpreted as the player $A$ being very inconsistent in their performance. However, there is no analogue to the luck function $\Lambda(x,y) = \thalf$, essentially because there is no uniform probability distribution on $\R$. We will see in \cref{subsection:practical_lambda} that this discrepancy is significant.
\end{example}

\begin{example}\label{example:BT}
  The \textit{Bradley--Terry model} \cite{zermelo, BT} is the special case of \cref{model:match} where $\cS = \R_{>0}$, $$\Lambda(x,y) = \frac{x}{x+y},$$ and $\mu_A$ and $\mu_B$ are Dirac measures. These choices can be interpreted as
  \begin{itemize}
    \item the players $A$ and $B$ always perform at the same strength, and
    \item it is possible for a match to be won by the player that played worse in that match. 
  \end{itemize}
  The Bradley--Terry model is often encountered under the reparameterizations $x \mapsto e^x$ \cite{glicko2} or $x \mapsto 10^{\frac{x}{400}}$ \cite{fide, glicko}.
\end{example}

\begin{example}\label{example:linear_models}
  ``Linear models'' in the sense described by David in \cite[\S 1.3, \S 4]{david} is the special case of \cref{model:match} where $\cS = \R$ and $\Lambda(x,y) = H(x-y)$. As discussed in \cite{david}, the Thurstone--Mosteller \cite{thurstone1, thurstone2, mosteller1, mosteller2, mosteller3} and Bradley--Terry \cite{zermelo, BT} models are recovered by requiring that $\mu_A$ and $\mu_B$ be normal distributions or Gumbel distributions with scale parameter $1$. In contrast with \cref{example:BT}, these choices can be interpreted as
  \begin{itemize}
  \item the players $A$ and $B$ perform at different strengths from one match to the next, and
  \item every match is won by whichever player played best.
  \end{itemize}
  Note that both this example and \cref{example:BT} recover the Bradley--Terry model, but do so in different ways, and can be interpreted differently.
\end{example}

\begin{example}\label{example:trueskill}
  The TrueSkill system \cite{trueskill, trueskill2} uses the special case of \cref{model:match} in which $\cS = \R$,
  \begin{align*}
    \Lambda(x,y) = \begin{cases} 1 & x-y > \eps \\ \half & |x-y| \leq \eps \\ 0 & x-y < -\eps\end{cases}
  \end{align*}
  for some $\eps > 0$, and $\mu_A$ and $\mu_B$ are normal distributions.
\end{example}

\begin{example}\label{example:fide}
  The rating system \cite[\S 8]{fide} used by FIDE, the de facto governing body of chess, corresponds to \cref{model:match} with $\cS = \R$, a choice of $\Lambda$ which is piecewise constant but well-approximated by the modified Bradley--Terry model
  \begin{align*}
    \Lambda(x,y) =
    \begin{cases}
      \frac{10}{11} & \text{if $x-y > 400$} \\
      \frac{1}{11} & \text{if $x-y < -400$}\\
      \frac{1}{1 + 10^{\frac{y-x}{400}}} & \text{if $|x-y| \leq 400$},
    \end{cases}
  \end{align*}
  and $\mu_A, \mu_B$ Dirac measures.
\end{example}

\begin{example}\label{example:randochess}
  Consider the game $G_\beta$ in which a chess match is played with probability $\beta \in (0,1)$, and a winner is chosen uniformly at random with probability $1 - \beta$. If the game of chess is perfectly modeled by the Bradley--Terry model as presented in \cref{example:BT}, then the game $G_\beta$ is best modeled by the choice of parameters $\cS = \R_{>0}$, $$\Lambda(x,y) = \beta\frac{x}{x+y} + (1-\beta)\half,$$ and $\mu_A$ and $\mu_B$ Dirac measures.
  
  None of the previous examples are good models for games in which it is impossible to win with probability arbitrarily close to $1$. For instance, they cannot properly model the fact that the world champion of chess would win a match of $G_\beta$ against both the author and a rock with probability $\tfrac{1 + \beta}{2}$, and the author would also win against the rock with probability $\tfrac{1 + \beta}{2}$. These sorts of considerations also cause problems when using the Rasch model \cite{rasch} for multiple-choice tests where guessing is possible \cite{HL}.
\end{example}

\begin{example}\label{example:blotto}
  Take $\cS = \R^3$ and
  \begin{align*}
    \Lambda\big((x_1,x_2,x_3), (y_1, y_2, y_3)\big)
    = 
    \begin{cases}
      1 & \text{if $\#\{i \in \{1,2,3\}\,:\, x_i > y_i\} \geq 2$}\\
      0 & \text{if $\#\{i \in \{1,2,3\}\,:\, y_i > x_i\} \geq 2$}\\
      \half & \text{otherwise}.
    \end{cases}
  \end{align*}
  For Dirac measures $\mu_A, \mu_B, \mu_C \in \cP(\cS)$ defined by
  \begin{align*}
    \mu_A(\{(3,3,3)\}) = 1,\quad \mu_B(\{(4,4,1)\}) = 1, \quad \mu_A(\{(5,2,2)\}) = 1,
  \end{align*}
  we have
  \begin{align*}
    L(\mu_A, \mu_B) = L(\mu_B, \mu_C) = L(\mu_C, \mu_A) = 0.
  \end{align*}
  This sort of non-transitivity arises in human play and is also an obstacle for machine learning \cite{CGTTOBJ, SWY, YDRWZC}. It is unclear how one would properly model this non-transitivity with the widely-used linear models that were presented in \cref{example:linear_models}.
\end{example}
\cref{model:match} makes two assumptions about the function $\Lambda$ which we use to simplify \cref{model:update}. However, there are situations in which these assumptions aren't appropriate. We discuss the assumptions below. In both cases, it seems straightforward conceptually to omit the assumption, but then one must repeat the work done in subsequent sections without using the associated simplifications.

The first assumption is that $\Lambda(x,y) = 1 - \Lambda(y,x)$. This is appropriate for symmetric games, but probably isn't for asymmetric games. For example, the game in which players flip a coin to determine who plays with white in a game of chess is symmetric, but the game of chess after having picked colours is not. At the time of writing, there have been $3,\!988,\!065,\!350$ matches of chess played on \url{lichess.org}, and in them white scored $52\%$ \cite{lichess}. To model this, one might instead fix two different $\Lambda$'s, each satisfying $\Lambda(x,x) = 0.5 \pm 0.02$, and consider which player was playing with the white pieces to determine which one to use.

The second assumption, that $\Lambda(x,y) \in [0,1]$, builds on the first. There are situations where one is interested not in the probability of each player winning, but other notions of score. For example, in poker ``cash games'' players can exchange money for chips at any time, and strive to win as many chips as possible. Thus, if $A$ and $B$ win $a$ and $b$ chips respectively while playing a prescribed number of hands against each other in a poker cash game, then it is most meaningful to consider the quantities $a$ and $b$ themselves, and not whether or not $a > b$. This contrasts with games like Go, where the winner of the match is the player with the highest score, regardless of the magnitude of the scores or the difference between them. It seems more natural to model poker cash games with a choice of $\Lambda$ which takes values outside of $[0,1]$. This also allows one to consider games which are not zero-sum. In two player games, when one player wins, the other loses, but other notions of score needn't sum to zero. In the preceding poker example, one expects that $a + b < 0$, as casinos take a small portion of each pot.

\subsection{Knowledge of players}\label{section:knowledge_of_players}
\cref{model:match} posits that the player $A$ is determined by a Borel probability measure $\mu_A \in \cP(\cS)$. Our rating system will not know with certainty what the true underlying measure $\mu_A$ is, and represents its current beliefs with a Borel probability measure $\nu_A \in \cP(\cP(\cS))$. The only observations our rating system will consider are match outcomes, and \cref{model:update} gives the resulting posteriors.
\begin{model}\label{model:update}
  Let $\cS$ and $L$ be as in \cref{model:match}. For any $\nu_A$ and $\nu_B$ in $\cP(\cP(\cS))$, define $\nu_{A, A>B}$ to be any element of $\cP(\cP(\cS))$ satisfying, for all Borel-measurable sets $U \subseteq \cP(\cS)$,
  \begin{align*}
    \int_U \nu_{A, A>B}(d\mu) = \frac{\int_U \left[\int_{\cP(\cS)} L(\mu,\mu') \,\nu_B(d\mu') \right] \nu_A(d\mu)}{\int_{\cP(\cS)} \left[\int_{\cP(\cS)} L(\mu,\mu') \,\nu_B(d\mu') \right] \nu_A(d\mu)}.
  \end{align*}
  Define $\nu_{A, B>A}$ as above but with all instances of $L(\mu,\mu')$ replaced by $L(\mu',\mu)$.
  
  The prior distribution over probability measures associated to a player $A$ is modeled by $\nu_A$, and the posterior after observing a match in which $A$ defeats $B$ or $B$ defeats $A$ is modeled by $\nu_{A, A>B}$ or $\nu_{A, B>A}$.
\end{model}
After updating the prior $\nu_A$ to the posterior $\nu_{A,A>B}$ or $\nu_{A,B>A}$, that posterior will then be used as the prior when processing the next match $A$ plays.

In some games there are match outcomes such as draws which are neither wins nor losses. When one can reasonably model these outcomes by a real number $\theta$, e.g.\ $\theta = \thalf$ for draws, then we suggest taking the posterior, which we'll denote $\nu_{A,\theta}$, to be any probability measure which satisfies
\begin{align}\label{model:update_theta}
  \int_U \nu_{A, \theta}(d\mu) = \frac{\int_U \left[\int_{\cP(\cS)} L(\mu,\mu')^\theta L(\mu',\mu)^{1-\theta} \,\nu_B(d\mu') \right] \nu_A(d\mu)}{\int_{\cP(\cS)} \left[\int_{\cP(\cS)} L(\mu,\mu')^\theta L(\mu',\mu)^{1-\theta} \,\nu_B(d\mu') \right] \nu_A(d\mu)}
\end{align}
for all Borel-measurable sets $U \subseteq \cP(\cS)$. Our reasoning for this is the same as the reasoning given in \cite[\S 2]{glicko}. This formula can also be viewed as encompassing the one given in \cref{model:update} if one takes $\theta = 1$ if $A>B$ and $\theta = 0$ if $B>A$. It may be helpful in some cases to note that the assumptions in \cref{model:match} imply that $L(\mu',\mu) = 1 - L(\mu,\mu')$.

One must choose what prior $\nu_0 \in \cP(\cP(\cS))$ to assign to a player which is completely unknown to the system. Below, we give one type of prior which is a convenient choice for many applications.
\begin{definition}\label{def:dirac_only}
  For a given complete separable metric space $\cS$, we will call a Borel probability measure $\nu \in \cP(\cP(\cS))$ \textit{Dirac-only} if and only if, for all Borel-measurable $U \subseteq \cP(\cS)$,
  \begin{align*}
    U \cap \big\{\delta_x \,:\, x \in \cS\big\} = \emptyset \quad\implies\quad \nu(U) = 0,
  \end{align*}
  where $\delta_x$ is the Dirac measure as in \cref{def:dirac_measure}.
\end{definition}
\noindent
Dirac-only priors are convenient for two reasons:
\begin{enumerate}
  \item If $\nu_A$ is Dirac-only, then the posteriors $\nu_{A,A>B}$ and $\nu_{A,B>A}$ from \cref{model:update} are too.
  \item If $\nu$ is Dirac-only, then we can treat $\nu$ as an element of $\cP(\cS)$ via
    \begin{align*}
      \nu(U \subseteq \cS) \coloneqq \nu\!\left(\big\{ \delta_x \,:\, x \in U\big\}\right).
    \end{align*}
\end{enumerate}

\begin{example}\label{example:glicko_update}
  This example gives a description of the Bayesian inference part of the widely-used Glicko system \cite{glicko}. Take $\cS = \R$ and $\Lambda$ to be the parameterization of the Bradley--Terry model \cite{zermelo, BT} given by
  \begin{align*}
    \Lambda(x,y) = \frac{1}{1 + 10^{\frac{y-x}{400}}}.
  \end{align*}
  To an unknown player, assign a Dirac-only (\cref{def:dirac_only}) prior $\nu_0$ which, viewed as a probability measure on $\R$, is a normal distribution. When updating according to \cref{model:update}, approximate the marginal likelihood $\int_{\cP(\cS)} L(\mu,\mu')\,\nu_B(d\mu')$ by a normal density with the same mode and second derivative at that mode; see \cite[Appendix A]{glicko} for details. A consequence of this approximation is that the posterior $\nu_{A,\theta}$ (\cref{model:update_theta}) is again normal.
\end{example}

\begin{example}\label{example:aliceabi_update}
  Take $\cS = \R$ and $\Lambda(x,y) = H(x-y)$, where $H$ is the Heaviside function (\cref{def:heaviside}). Suppose Alice and Abi share the account $A$ in an online game, and in any given match Alice plays with probability $p \in (0,1)$ and Abi plays otherwise. When Alice plays she always performs with strength $x_1$, and when Abi plays she always performs with strength $x_2$. Suppose there is another player $B$ which always performs with strength $y$.
  
  In the case where our rating system is aware of all this information except for the value of $p$, we can model the situation as follows. Let $\mu_p$ be a probability measure satisfying $\mu_p(\{x_1\}) = p$ and $\mu_p(\{x_2\}) = 1-p$. A reasonable choice of prior $\nu_A$ would be the measure which, for any Borel-measurable subset $U$ of $[0,1]$, gives probability $\lambda(U)$ to the set $\{\mu_p \,:\, p \in U\}$, where $\lambda$ is the Lebesgue measure. This corresponds to a uniform prior for $p$. The prior $\nu_B$ should be taken to be $\delta_{\delta_y}$ (\cref{def:dirac_measure}).
  
  If $x_1 < y < x_2$, then observations of match outcomes between $A$ and $B$ are essentially samples of Bernoulli random variable with unknown parameter $p$, and the rating system is attempting to determine $p$. \cref{model:update} uses the usual Bayesian inference to estimate $p$ and yields a beta distribution.
  
  If one considers only Dirac-only choices of $\nu_A$, like in \cref{example:glicko_update}, then one cannot reasonably model the situation given in this example. If it is known that $B$ always performs with strength $y$ and that every match is won by the player who performed better, then, after a single observation of $A>B$, the posterior $\nu_{A,A>B}$ gives probability $0$ to all $\delta_x$ with $x < y$. The posterior after observing both $A>B$ and $B>A$ would give probability $0$ to all $\delta_x$ with $x \neq y$. If $p$ is far from $\thalf$, then $\delta_y$ would be a very poor guess for $\mu_A$. If there is another player $C$ who is known to always perform with strength $z$ satisfying $x_1 < y < z < x_2$, then assuming $\mu_A$ is a Dirac measure would cause system to believe that the sequence of match outcomes $A>B$, $B>A$, and $A>C$ could never occur, and in the proportion $p(1-p)^2$ of cases in which it does the system's behaviour would be undefined.
\end{example}

\begin{example}\label{example:discrete_update}
  In this example we restate \cref{model:update} under the assumption that $\nu_A$ and $\nu_B$ are discrete distributions over discrete distributions. We use the following notation:
  \begin{itemize}
    \item $(\alpha_i)_i$ and $(\beta_j)_j$ are integer-indexed sequences of non-negative real numbers which each sum to $1$, as are $(p_{i,k})_k$ for each $i$, and $(q_{j,\ell})_\ell$ for each $j$.
    \item $(x_{i,k})_k$ and $(y_{j,\ell})_\ell$ are integer-indexed sequences of elements of $\cS$.
    \item $\delta_*$ is the Dirac measure at $*$ (\cref{def:dirac_measure}).
    \item $\propto$ means that values should be normalized so that they sum to $1$.
  \end{itemize}
  Write
  \begin{alignat*}{2}
    \nu_A &= \sum_i \alpha_i \delta_{\mu_{A,i}},\quad &\nu_B &= \sum_j \beta_j \delta_{\mu_{B,j}},\\
    \mu_{A,i} &= \sum_k p_{i,k} \delta_{x_{i,k}},\quad &\mu_{B,j} &= \sum_\ell q_{j,\ell} \delta_{y_{j,\ell}}.
  \end{alignat*}
  Then
  \begin{align*}
    L(\mu_{A,i},\mu_{B,j}) = \sum_{k,\ell} p_{i,k}q_{j,\ell} \Lambda(x_{i,k}, y_{j,\ell})
  \end{align*}
  and
  \begin{align*}
    &\nu_{A,A>B} = \sum_i \hat\alpha_i \delta_{\mu_{A,i}}, \quad\quad \hat\alpha_i \propto \alpha_i \sum_j \beta_j \sum_{k,\ell} p_{i,k}q_{j,\ell} \Lambda(x_{i,k}, y_{j,\ell}).\qedhere
  \end{align*}
\end{example}

\subsection{Player growth}\label{section:player_evolution}
In practice, the strength of a player $A$ is likely change over time for a variety of reasons. For example, $A$ might read a book to learn a new chess opening, or acquire new cards in a collectible card game. Using \cref{model:update} to process many of $A$'s matches can easily lead to situations where $\nu_A$ gives probability nearly $0$ to certain measures. If external factors then cause $A$'s strength to change, a purely Bayesian system might need to observe many match outcomes before it gives a non-negligible probability to the measure corresponding to $A$'s new strength. In this section, we present \cref{model:evolution} for how a player's underlying measure can change between matches.
\begin{model}\label{model:evolution}
  Let $\cS$ be a complete separable metric space. 
  Fix a function $\kappa: \cP(\cS) \to \cP(\cP(\cS))$ and denote by $\kappa_\mu$ the value of $\kappa$ evaluated at $\mu$. Given any $\nu \in \cP(\cP(\cS))$, define $\tilde{\nu}$ to be any element of $\cP(\cP(\cS))$ satisfying, for all Borel-measurable sets $U \subseteq \cP(\cS)$,
  \begin{align*}
    \int_U \tilde{\nu}(d\mu') \coloneqq \int_{\cP(\cS)} \int_U\kappa_\mu(d\mu') \,\nu(d\mu).
  \end{align*}
  
  The possibility of the strength of $A$ changing between matches because of external factors is modeled by replacing the measure $\nu_A$ \cref{model:update} associates to them by $\tilde{\nu}_A$.
\end{model}
We call the function $\kappa$ a \textit{kernel}.
\begin{example}\label{example:dirac_kernel}
  If $\kappa_\mu = \delta_\mu$ for all $\mu \in \cP(\cS)$, then $\tilde{\nu} = \nu$ for all $\nu$. Here $\delta_\mu$ denotes the Dirac measure at $\mu$ (\cref{def:dirac_measure}).
\end{example}

\begin{example}\label{example:glicko_kernel}
  Recall the description of the Glicko system given in \cref{example:glicko_update}. 
  Let $\varphi(\,\cdot\,|\,x,\sigma^2)$ be the normal density on $\R$ with mean $x$ and variance $\sigma^2$. As described in \cite[\S 3.2]{glicko}, one step of the Glicko system is using \cref{model:evolution} with $\kappa_{\delta_x}$ the Dirac-only measure (\cref{def:dirac_only}) satisfying
  \begin{align*}
    \kappa_{\delta_x}(\{\delta_y\,:\, y\in U\}) = \int_U \varphi(y\,|\,x,\sigma^2)\,dy
  \end{align*}
  for some fixed $\sigma^2$ and all Borel-measurable $U \subseteq \R$. As mentioned in \cref{example:glicko_update}, this formulation of Glicko only involves $\nu$ which are Dirac-only, so the value $\kappa_\mu$ when $\mu$ isn't a Dirac measure is irrelevant. It's computationally convenient that if $\nu$ and $\kappa$ are normal in this sense, then $\tilde{\nu}$ is again normal; c.f.\ \cite[Eq.\ (7)]{glicko}.
\end{example}

\begin{remark}\label{remark:glicko2}
  In some applications allowing $\kappa$ to depend on the player can increase the accuracy of the model. The main innovation of Glicko2 \cite{glicko2} over Glicko is choosing such a dependence. However, this can incentivize players seeking to maximize their rating to intentionally lose games in some circumstances \cite[\S 5.1]{EL}. Competitive players of Pok\'{e}mon GO have done this and perceive the resulting rankings to be inaccurate \cite{thesilphroad}.
\end{remark}

\section{Parameter choices}\label{section:practical}
Our system is in use in the online collectible card game Duelyst \rtwo. To implement the system, we needed to choose values for the parameters listed at the beginning of \cref{section:model}:
\begin{itemize}
  \item In \cref{model:match}, a luck function $\Lambda$
  \item In \cref{model:update}, a prior $\nu_0$ for unknown players
  \item In \cref{model:evolution}, a kernel $\kappa$
\end{itemize}
In this section we present the choices we made and give a concrete description of the resulting system. Our system is succinctly summarized in \cref{subsection:practical_summary}. We expect that similar choices will be suitable for many applications, and discuss minor variations other situations might call for.

\subsection{The luck function $\Lambda$ and its tails}\label{subsection:practical_lambda}
For Duelyst \rtwo, we take $\cS = \R$. There are situations like those discussed in \cref{example:blotto} which call for different choices of $\cS$, but we didn't feel that this was necessary for our application.

Our choice of $\Lambda$ is
\begin{align}\label{eq:d2_lambda}
  \Lambda(x,y) = \frac{1-\beta}{2} + \frac{\beta}{1 + \exp(y-x)}
\end{align}
with $\beta = 0.8$. This is the $\Lambda$ from \cref{example:randochess}, and we discuss it there. When displaying ratings in game, we first apply the transformation
\begin{align}\label{eq:d2_parameterization}
  x \mapsto \frac{400}{\log 10}x + 1500
\end{align}
so that we match the parameterization of the Bradley--Terry model used by FIDE \cite{fide} and Glicko \cite{glicko}.

Our $\Lambda$ is a linear combination of a constant function and a sigmoid. We chose a logistic curve for the sigmoid, but we would guess that other choices, like a normal CDF as in the Thurstone--Mosteller model \cite{thurstone1, thurstone2, mosteller1, mosteller2, mosteller3}, would work just as well. In contrast, the constant term in $\Lambda$ has a very large impact on the behaviour of the system.
\begin{figure}[H]
  \includegraphics[width=\textwidth]{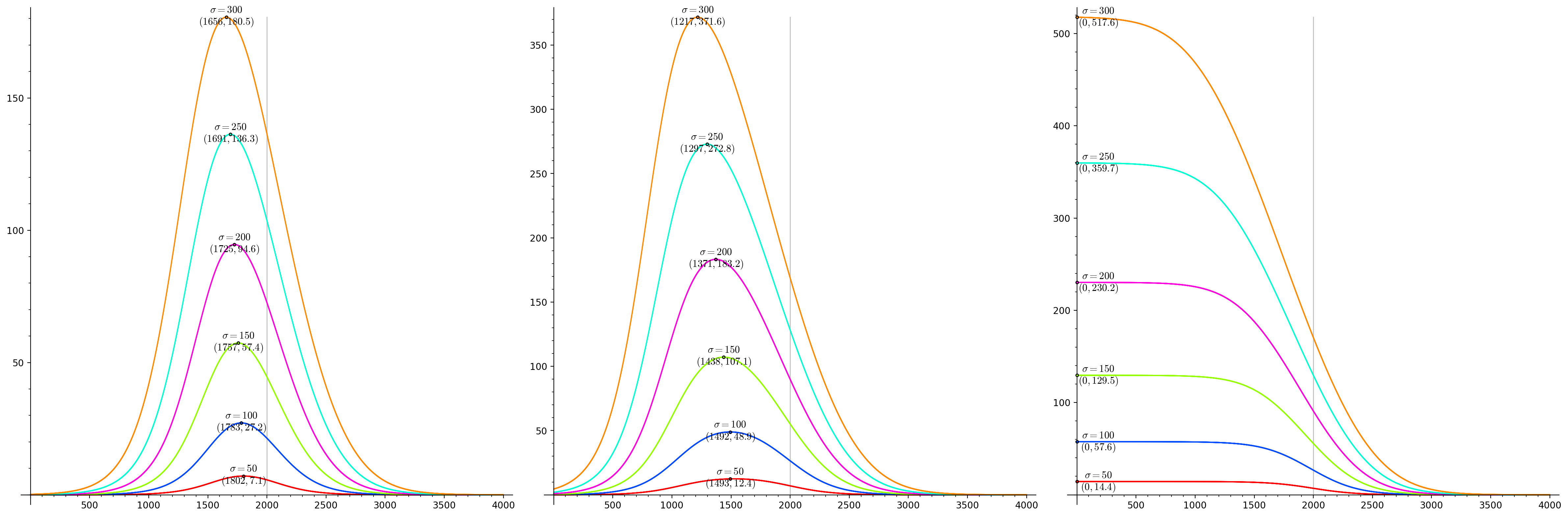}
\end{figure}
\begin{figurecap}\label{fig:delta_mu}
  Difference in means \eqref{eq:delta_mu} between $\nu_A \sim \cN(m,\sigma^2)$ and $\nu_{A,A>B}$ as a function of $m$. Here $\nu_B = \delta_{\delta_{2000}}$, and $\Lambda$ is given by \eqref{eq:d2_lambda} and \eqref{eq:d2_parameterization} with, from left to right, $\beta = 0.8$, $0.99$, and $1$. The maximum differences for $m \in [0,4000]$ are highlighted.
\end{figurecap}
\cref{fig:delta_mu} illustrates the effect of including a constant term in $\Lambda$. We suppose $B$'s strength is always $2000$ (i.e.\ $\nu_B = \delta_{\delta_{2000}}$; c.f.\ \cref{def:dirac_measure}), $A$'s strength is an unknown real number (i.e.\ $\nu_A$ is ``Dirac-only'' as in \cref{def:dirac_only}), and the system's prior $\nu_A$ for $A$'s strength is a normal distribution with mean $m$ and variance $\sigma^2$. If the match outcome $A > B$ is observed, then the system uses \cref{model:update} to update its belief about $A$'s strength. Let $m'$ be the mean of $A$'s updated distribution $\nu_{A,A>B}$. \cref{fig:delta_mu} plots the difference $m' - m$ as a function of $m$ for $\Lambda$ as in \eqref{eq:d2_lambda} and reparameterized according to \eqref{eq:d2_parameterization}, $\beta = 0.8$, $0.99$, $1$, and various $\sigma$. Unpacking the definitions, one can write the quantity being plotted explicitly:
\begin{align}\label{eq:delta_mu}
  &m' - m = \frac{\ds{\frac{1-\beta}{2}m + \int_\R \frac{\beta}{1 + 10^{\frac{2000-x}{400}}}\exp\!\left(-\frac{(x-m)^2}{2\sigma^2}\right) \frac{x\,dx}{\sqrt{2\pi\sigma^2}}}}{\ds{\frac{1-\beta}{2} + \int_\R \frac{\beta}{1 + 10^{\frac{2000-x}{400}}}\exp\!\left(-\frac{(x-m)^2}{2\sigma^2}\right) \frac{dx}{\sqrt{2\pi\sigma^2}}}} - m.
\end{align}

When $\beta < 1$, one interpretation is that, like in \cref{example:randochess}, there is a nonzero chance that the winner of a match is decided uniformly at random. If $B$ is vastly stronger than $A$, then, when the match outcome $A > B$ is observed, the only plausible explanation is that the outcome of the match was in fact decided by chance, and the posterior distribution $\nu_{A,A>B}$ for $A$ is very similar to their prior $\nu_A$. In \eqref{eq:delta_mu}, this intuition is reflected in the value of the integrals being much smaller than the constant terms, because there is nearly no overlap between the factors
$$\frac{\beta}{1 + 10^{\frac{2000-x}{400}}} \quad\text{and}\quad \exp\!\left(-\tfrac{(x-m)^2}{2\sigma^2}\right),$$ which respectively come from the sigmoidal term in $\Lambda$ (defined in \eqref{eq:d2_lambda}) and the prior $\nu_A$.

In contrast, when $\beta = 1$, the constant terms in the numerator and denominator of \eqref{eq:delta_mu} vanish, and the difference in the means of $\nu_A$ and $\nu_{A,A>B}$ is the ratio of the two exponentially small integrals. Essentially, the system views the chance of $A$ having strength comparable to $B$'s as vanishingly small, but also views the chance of observing the match outcome $A > B$ as vanishingly small. Because $\nu_A$ has the $\exp(-x^2)$ tails of a normal distribution, but $\Lambda$ has the much heavier tail $\exp(-x)$, almost all of the mass in the integrals comes from $x \approx m$. A straightforward calculation shows that $$m' - m \longrightarrow \frac{\sigma^2\log 10}{400}$$ as $m \longrightarrow -\infty$.

The behaviour with $\beta = 1$ overall seems much less reasonable to us than when $\beta < 1$, since it leads to massive rating changes for $A$ based on ratios of minuscule probabilities. We view these probabilities as being substantially smaller than the chance that something bizarre has happened that the $\beta = 1$ model is not equipped to consider, and believe that a model that better reflects reality should not view this match outcome as extremely strong evidence that $A$ is tremendously underrated. \cref{fig:delta_mu} shows that changing $\beta$ from $1$ to $0.99$ changes the behaviour of the model much more than changing $\beta$ from $0.99$ to $0.8$.

\begin{wrapfigure}{r}{0.42\textwidth}
  \vspace{-1.3\baselineskip}
  \includegraphics[width=0.42\textwidth]{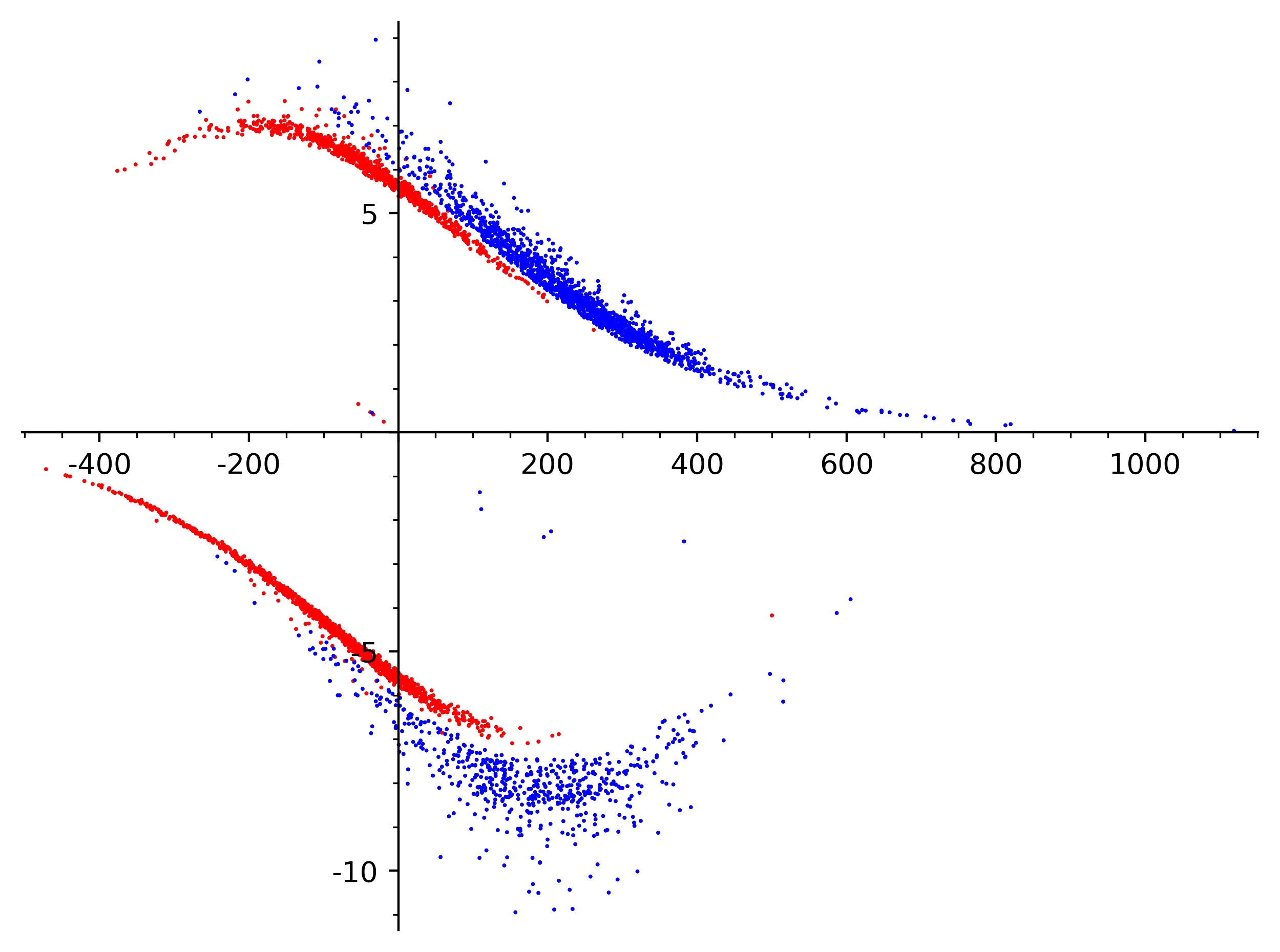}
\begin{figurecap}\label{fig:delta_mu_empirical}
  Change in rating after each match for the two Duelyst \rtwo\hspace{0cm} players that have played the most matches at the time of writing, plotted against difference between their rating and their opponent's. The first $100$ matches from each player are excluded.
\end{figurecap}
\vspace{-\baselineskip}
\end{wrapfigure}

We note that \cite[Fig.\ 1]{ghb} reports that four widely-used systems based on the Bradley--Terry model all overestimate the performance of very highly rated competitors, that Sonas reports observing the same phenomenon in the popular article \cite{sonas} 
with a dataset of $1.54$ million chess games from FIDE
, and that FIDE truncates the tails of the $\Lambda$ it uses; see \cref{example:fide}. This is consistent with our qualitative assessment of \cref{fig:delta_mu} above: the Bradley--Terry model, with exponential asymptotes at $0$ and $1$, overestimates the winning chances of a much higher rated player.

The two players who have played the most matches in our Duelyst \rtwo\hspace{0cm} dataset of the first $1,\!126,\!592$ ranked matches played since the game's launch, who we call $P$ and $Q$. have played $2162$ and $2142$ matches respectively, and are ranked by our system to be at the $95^{\text{th}}$ and $99.95^{\text{th}}$ percentiles among all players having played at least one ranked match. \cref{fig:delta_mu_empirical} shows the change in rating, i.e.\ mean of $\nu_P$ or $\nu_Q$, after each of their matches beyond the first $100$ they played. The horizontal axis shows the rating difference at the time of the match between the player and their opponent. $P$ is shown in red, and $Q$ in blue. Points above the axis are wins and points below are losses, except for the $10$ distinctly visible draws separate from the rest of the points.%

The shape of the clusters in \cref{fig:delta_mu_empirical} is similar to what's shown in \cref{fig:delta_mu}, as expected. The qualitative observation that the blue points are noisier than the red points is explained by the fact that $\text{Var}(\nu_Q)$ varies between about $52^2$ and $62^2$ for the plotted matches, whereas $\text{Var}(\nu_P)$ is almost always between $48^2$ and $52^2$.

\subsection{$\nu_0$ for unknown players}\label{subsection:practical_nu}
Duelyst \rtwo's implementation considers only values of $\nu$ which are Dirac-only (\cref{def:dirac_only}). We can then view $\nu$ as a probability distribution on $\R$, with the interpretation that each player's strength could in principle be described by a single real number, and $\nu$ describes the system's knowledge of said real number. The prior $\nu_0$ we assign to an unknown player, viewed as a distribution on $\R$, is
\begin{align}\label{eq:d2_nu}
  \nu_0 = \sum_{k=0}^{n} \rho(x_k) \,\delta_{x_k}
\end{align}
with
\begin{align*}
  &x_k = -M + \frac{2M}{n}k,
  \quad\text{}\quad
  \rho(x_k) \propto \varphi(x_k \,|\, 0, \sigma_0^2),
\end{align*}
\vspace{-\baselineskip}
\begin{align*}
  &n = 1000,
  \quad\text{}\quad
  M = 7,
  \quad\text{}\quad
  \sigma_0^2 = 0.7^2.
\end{align*}

Here $\delta_{x_k}$ is a Dirac measure (\cref{def:dirac_measure}), $\varphi(\,\cdot \,|\, x,\sigma_0^2)$ is the normal density with mean $x$ and variance $\sigma_0^2$, and $\propto$ indicates that the values of $\rho(x_k)$ are rescaled so that they sum to $1$.

It would be more natural to take $\nu_0$ to be the normal distribution $\cN(0,0.7^2)$, but we choose the discrete distribution \eqref{eq:d2_nu} with finite support because it is straightforward to implement: we can encode \eqref{eq:d2_nu} as the tuple $\left(\rho(x_0), \dots, \rho(x_{1000})\right) \in \R^{1001}$, a list of $1001$ real numbers. The measure \eqref{eq:d2_nu} is an approximation to $\cN(0,0.7^2)$ in the sense of weak convergence as $n, M \longrightarrow \infty$.
 
We chose the values of $n=1000$, $M=7$, and $\sigma_0^2 = 0.7^2$ by examining the system's performance on the dataset of the first $1,\!126,\!592$ ranked matches played since Duelyst \rtwo's launch. We were primarily concerned with producing a reliable ranking for the game's strongest players. The resulting rankings were relatively insensitive to these choices.

We found that, for this dataset, taking $\sigma_0^2 = 1$ lead to a small but non-negligible chance for players who happened to do very well in their first few matches getting ranked inappropriately highly, whereas taking $\sigma_0^2 = 0.7^2$ seemingly did not. Taking smaller values of $\sigma_0^2$ would require top players to play more matches to be accurately rated.

The numbers $M$ and $n$ control how precisely the system can determine a player's strength, but larger values make the system more computationally expensive to use in practice. We chose $M = 7$ because no player had non-negligible mass outside $[-M,M]$, and $n=1000$ because doubling this value did not meaningfully change the system's output for our dataset.

\subsection{The kernel $\kappa$}\label{subsection:practical_kernel}
For Duelyst \rtwo, we took $\kappa$ to be
\begin{align}\label{eq:d2_kappa}
  \kappa_{\delta_x} = \sum_{k=0}^{n} K(x,x_k)\,\delta_{x_k}
\end{align}
with
\begin{align*}
  &x_k = -M + \frac{2M}{n}k,
  \quad\text{}\quad
  K(x, x_k) \propto \varphi(x \,|\, x_k, \sigma_\kappa^2),
\end{align*}
\vspace{-\baselineskip}
\begin{align*}
  &n = 1000,
  \quad\text{}\quad
  M = 7,
  \quad\text{}\quad
  \sigma_\kappa^2 = 0.03^2,
\end{align*}
using the same notation as \cref{eq:d2_nu}. Because all $\nu$'s that arise will be Dirac-only (\cref{def:dirac_only}), the value of $\kappa_\mu$ when $\mu$ is not a Dirac measure irrelevant. As was the case in \cref{subsection:practical_nu}, a more natural choice for $\kappa$, viewed as a distribution on $\R$, would be the normal distribution $\cN(x_k, \sigma_\kappa^2)$ (and in some applications it might be desirable to take $\kappa$ to be, more generally, a mixture of normal distributions with the same mean), but, for computational convenience, we want $\tilde{\nu}$ to give $0$ mass to $\R \backslash \{x_0, \dots, x_n\}$.

\subsection{Summary}\label{subsection:practical_summary}
In Duelyst \rtwo, each player $A$ is represented as a tuple of $1001$ real numbers:
$$\bigg(\rho_A(\tfrac{14}{1000}\cdot 0 - 7),\, \rho_A(\tfrac{14}{1000}\cdot 1 - 7), \dots,\, \rho_A(\tfrac{14}{1000}k - 7), \dots,\, \rho_A(7)\bigg).$$
New players are set such that
$$\rho_A(x) \propto \frac{1}{\sqrt{2\pi\cdot 0.7^2}}\exp\!\left(-\frac{x^2}{2\cdot 0.7^2}\right).$$
Here and later, $\propto$ indicates that the values are then normalized so that they sum to $1$.

After the match outcome $A > B$ is observed, the system updates the values of $\rho_A$ and $\rho_B$ to
\begin{align*}
  &\rho_{A,A>B}(x) \propto \rho_A(x) \sum_{k=0}^{1000} \rho_B\!\left(\tfrac{14}{1000}k - 7\right)\left[0.1 + \frac{0.8}{1 + \exp\!\left((\frac{14}{1000}k - 7) - x\right)}\right],\\
  &\rho_{B,A>B}(x) \propto \rho_B(x) \sum_{k=0}^{1000} \rho_A\!\left(\tfrac{14}{1000}k - 7\right)\left[0.1 + \frac{0.8}{1 + \exp\!\left(x - (\frac{14}{1000}k - 7)\right)}\right].
\end{align*}
We evaluate these expressions using the Fast Fourier Transform (FFT) as described in \cref{subsection:algorithms_fft}. The values of $\rho_A$ and $\rho_B$ are then replaced with the values of $\rho_{A,A>B}$ and $\rho_{B,A>B}$.

Immediately after completing the step above, $\rho_A$ is replaced with $\tilde{\rho}_A$, defined by
\begin{align*}
  \tilde{\rho}_A(x) \propto \sum_{k=0}^{1000} \rho_A\!\left(\tfrac{14}{1000}k - 7\right) \frac{1}{\sqrt{2\pi \cdot 0.03^2}} \exp\!\left(-\frac{\left(x - (\tfrac{14}{1000}k - 7)\right)^2}{2\cdot 0.03^2}\right),
\end{align*}
and $\rho_B$ is replaced with $\tilde{\rho}_B$ defined analogously. FFT is used to compute the values of $\tilde{\rho}_A$ and $\tilde{\rho}_B$.

\section{Algorithms}\label{section:algorithms}
Our rating system operates as follows:
\begin{itemize}
  \item Assign a prior $\nu_0$ to unknown players (see \cref{model:update} and \cref{subsection:practical_nu}).
  \item After every match:
    \begin{enumerate}
    \item update $\nu_A$ and $\nu_B$ to $\nu_{A,A>B}$ and $\nu_{B, A>B}$ using to \cref{model:update},
    \item replace $\nu_A$ and $\nu_B$ with $\tilde{\nu}_A$ and $\tilde{\nu}_B$ using \cref{model:evolution},
    \end{enumerate}
  where $A$ and $B$ are the winner and loser of that match respectively.
\end{itemize}
We call step 1 above \textit{match processing}, and step 2 \textit{kernel processing}.

In this section, we present three algorithms for each of these steps, for three special cases of parameter choices. In all three cases, we will assume that the playing strength of an arbitrary player $A$ is an unknown fixed real number that is an element of a known finite set $S_A$ (which can depend on $A$) of size at most $n+1$ (which cannot depend on $A$).

In \cref{subsection:algorithms_naive}, we give an algorithm that is fully general besides the assumptions above. This algorithm takes time $\gg n^2$ for each of the match processing step from \cref{model:update} and the kernel processing step from \cref{model:evolution}, whereas the other two algorithms we present take time $\ll n^{1+\eps}$ for these steps. This algorithm is useful because it is the simplest of the three.

In \cref{subsection:algorithms_fft}, we give an algorithm based on the Fast Fourier Transform (FFT) that processes matches and kernels in time $\ll n^{1+\eps}$, but requires some mild additional assumptions. 
This is the algorithm that we think is most useful for applications, and that we used in Duelyst \rtwo.

In \cref{subsection:algorithms_laplace}, we give another algorithm that processes matches and kernels in $\ll n^{1+\eps}$, but does not rely on FFT, and instead is completely elementary. It makes the somewhat restrictive assumption that the functions $\Lambda$ and $\kappa$ are essentially short linear combinations of CDFs or PDFs of Laplace distributions respectively, but omits assumptions on $S_A$ that were necessary for the FFT-based algorithms.

\subsection{Naive algorithms}\label{subsection:algorithms_naive}

Let $\rho_A$ be such that $\rho_A(x) = \nu_A(\{x\})$ for all $x$. If the match $A > B$ is observed, then $A$'s posterior distribution is given by
\begin{align}\label{eq:naive_posterior}
  \rho_{A,A>B}(x) \propto \rho_A(x) \sum_{x_k \in S_B} \rho_B(x_k) \,\Lambda(x, x_k).
\end{align}

Evaluating $\rho_{A,A>B}(x)$ for a specific $x$ by summing the right hand side above directly takes time $\cO(n)$. The function $\rho_A$ is supported on at most $n$ points, so overall the posterior can be computed in time $\cO(n^2)$. If $B > A$ is observed instead, then the same formula can be used with $\Lambda(x, x_k)$ replaced by $\Lambda(x_k, x) = 1 - \Lambda(x, x_k)$.

\begin{example}\label{example:naive_algorithm_simple_match}
  Suppose
  \begin{align*}
    &\ds{\Lambda(x,y) = \frac{x}{x+y}},\\
    &\rho_A(2) = \frac{9}{20}, \quad\quad \rho_A(5) = \frac{3}{20}, \quad\quad \rho_A(13) = \frac{8}{20},\\
    &\rho_B(3) = \frac{2}{11}, \quad\quad \rho_B(7) = \frac{4}{11}, \quad\quad \rho_B(11) = \frac{5}{11}.
  \end{align*}
  If $A>B$ is observed, then
  \begin{align*}
    \rho_{A,A>B}(2)
    &\propto \rho_A(2)\sum_{x_k \in \{3,7,11\}}\rho_B(x_k)\frac{2}{2+x_k} = \frac{719}{7150},\\
    \rho_{A,A>B}(5)
    &\propto \frac{43}{704},\\
    \rho_{A,A>B}(13)
    &\propto \frac{208}{825}.
  \end{align*}
  Normalizing gives
  \begin{align*}
    \rho_{A,A>B}(2) = \frac{69024}{284005} \approx 0.24, \quad\quad \rho_{A,A>B}(5) = \frac{41925}{284005} \approx 0.15, \quad\quad \rho_{A,A>B}(13) = \frac{173056}{284005} \approx 0.61.
  \end{align*}
  Similarly,
  \begin{align*}
    \rho_{B,A>B}(3) = \frac{74724}{284005} \approx 0.26, \quad\quad \rho_{B,A>B}(7) = \frac{105456}{284005} \approx 0.37, \quad\quad \rho_{B,A>B}(11) = \frac{103825}{284005} \approx 0.37.
  \end{align*}
  Note that we use $\rho_A$, not $\rho_{A,A>B}$, to update $\rho_B$.
\end{example}

Let $K:\R^2 \to \R$ be defined by
\begin{align*}
  \kappa_{\delta_x} = \sum_{x_k \in S_A} K(x,x_k)\,\delta_{x_k}
\end{align*}
as in \eqref{eq:d2_kappa}. Then $\tilde{\rho}_A$ can be computed as
\begin{align}\label{eq:naive_kernel}
  \tilde{\rho}_A(x) \propto \sum_{x_k \in S_A} \rho_A(x_k) \,K(x,x_k)
\end{align}
for $x \in S_A$. Like the case with match processing discussed above, computing $\tilde{\rho}_A$ this way takes time $\cO(n^2)$.

\begin{example}\label{example:naive_algorithm_simple_kernel}
  Suppose $S_A = \{1,2,\dots,100\}$ and
  \begin{align*}
    &K(x,y) = \begin{cases} \frac{1}{3} & \text{if }x-y \in \{-1,0,1\}  \\ 0 & \text{otherwise,} \end{cases}\quad\quad
    \rho_A(n) = \begin{cases} \frac{1}{10} & \text{if $n$ is a perfect square} \\ 0 & \text{otherwise.} \end{cases}
  \end{align*}
  Then
  \begin{align*}
    &\tilde{\rho}_A(n) = \begin{cases} \frac{1}{28} & n \in U \\ 0 & \text{otherwise,} \end{cases}\\
    \shortintertext{where}
    U &= \big\{n \in \Z \cap [1,100] \,:\, \text{for some $\delta \in \{-1,0,1\}$, $n+\delta \in [1,100] \text{ and } \sqrt{n+\delta} \in \Z$}\big\}\\
    &= \{1, 2, 3, 4, 5, 8, 9, 10, 15, 16, 17, 24, 25, 26, 35, 36, 37, 48, 49, 50, 63, 64, 65, 80, 81, 82, 99, 100\}.\qedhere
  \end{align*}
\end{example}

\subsection{FFT-based algorithms}\label{subsection:algorithms_fft}
We write $f(n) = \tilde{\cO}(g(n))$ to mean $f(n) = \cO(g(n)n^\eps)$ for all $\eps > 0$. Define $\rho_A, \rho_B$, and $K$ as in the previous section. We will recognize \eqref{eq:naive_posterior} and \eqref{eq:naive_kernel} as convolutions, and then use the Fast Fourier Transform (FFT) to compute them in time $\tilde{\cO}(n)$. In this section, we assume the following:
\begin{itemize}
\item $S_A = S_B$,
\item $S_A = \{x_0, \dots, x_n\}$ with $x_k = k\Delta$ for some $\Delta \in \R_{>0}$, 
\item $\Lambda(x,y) = \tfrac{1-\beta}{2} + \beta F(x-y)$ for some $\beta \in [0,1]$ and $F : \R \to [0,1]$ increasing with asymptotes at $0$ and $1$,
\item $K(x,y) = G(x-y)$ for some $G : \R \to \R$.
\end{itemize}

We begin by presenting an algorithm for kernel processing. With the assumptions above, \eqref{eq:naive_kernel} can be written as
\begin{align*}
  \tilde{\rho}_A(k\Delta) = \sum_{j=0}^n \rho_A(j\Delta) \,G\big((k-j)\Delta\big).
\end{align*}
We recognize this as a discrete convolution: 
$\tilde{\rho}_A = \rho_A * G$. 
We can then compute all the values in the list $\tilde{\rho}_A = \big(\tilde{\rho}_A(0), \tilde{\rho}_A(\Delta), \dots, \tilde{\rho}_A(n\Delta)\big)$ in time $\tilde{\cO}(n)$ by using FFT.

Our algorithm for computing the posterior \eqref{eq:naive_posterior} is similar, but involves some additional elementary manipulations. Define
\begin{align*}
  R(x) \coloneqq F(x) - H(x),
\end{align*}
where $H$ is the Heaviside function (\cref{def:heaviside}). Our assumptions imply that $R(x) \longrightarrow 0$ as $|x| \longrightarrow \infty$, and that
\begin{align*}
  &\rho_{A,A>B}(k\Delta) \propto \rho_A(k\Delta) \left(\frac{1-\beta}{2} + \beta\big(L_R(k\Delta) + L_H(k\Delta)\big)\right) \text{ and}\\
  &\rho_{A,A<B}(k\Delta) \propto \rho_A(k\Delta) \left(\frac{1+\beta}{2} - \beta\big(L_R(k\Delta) + L_H(k\Delta)\big)\right),
\end{align*}
where
\begin{align*}
  &L_R(k\Delta) \coloneqq \sum_{j=0}^n \rho_B(j\Delta) \,R\big((k-j)\Delta\big)\quad\text{and}\quad
  L_H(k\Delta) \coloneqq \sum_{j=0}^n \rho_B(j\Delta) \,H\big((k-j)\Delta\big).
\end{align*}

The function $L_R$ is a convolution: $L_R = \rho_B * R$. We compute all values in the list $L_R = \big(L_R(0), \dots, L_R(n\Delta)\big)$ in time $\tilde{\cO}(n)$ using FFT.

The following algorithm computes all values in the list $L_H = \big(L_H(0), \dots, L_H(n\Delta)\big)$ in time $\cO(n)$.
\begin{algorithm}\label{algorithm:L_H} Compute $L_H(k\Delta)$ for all $k \in \{0,1,\dots,n\}$.\\
\begin{algorithmic}
  \State $\Sigma \gets 0$
  \For{$0 \leq k \leq n$}
    \State $\Sigma \gets \Sigma + \half \rho_B(k\Delta)$ \hspace{1.5cm}(Because $H(0) = \thalf$) 
    \State $L_H(k\Delta) \gets \Sigma$
    \State $\Sigma \gets \Sigma + \half \rho_B(k\Delta)$
  \EndFor
\end{algorithmic}
\end{algorithm}

\subsection{Laplace algorithms}\label{subsection:algorithms_laplace}
Let $\rho_A$, $\rho_B$, and $K$ be as in \cref{subsection:algorithms_naive}. Define
\begin{align*}
  &f(x\,|\,b) \coloneqq \frac{1}{2b}\exp\!\left(-\frac{|x|}{b}\right) \quad\text{and}\quad
  F(x\,|\,b) \coloneqq \begin{cases}\tfrac{1}{2}\exp\!\left(-\frac{|x|}{b}\right) & x < 0\\ 1 - \tfrac{1}{2}\exp\!\left(-\frac{|x|}{b}\right) & x \geq 0.\end{cases}
\end{align*}
These are respectively the PDF and CDF of a Laplace distribution. Assume
\begin{itemize}
\item $\Lambda(x,y) = \frac{1-\beta}{2} + \beta\sum_{j=1}^\ell p_j F(x-y \,|\,a_j)$ for non-negative reals $p_j$ which sum to $1$,
\item $K(x,y) = \sum_{j=1}^\ell q_j f(x-y \,|\, b_j)$ for non-negative reals $q_j$ which sum to $1$.
\end{itemize}
With these assumptions, \eqref{eq:naive_posterior} and \eqref{eq:naive_kernel} become
\begin{align*}
  \rho_{A,A>B} &\propto \rho_a(x)\left[\frac{1-\beta}{2} + \beta\sum_{j=1}^\ell p_j \sum_{x_k \in S_B}\rho_B(x_k)F(x - x_k\,|\,a_j)\right]\\
  \shortintertext{and}
  \tilde{\rho}_A(x) &\propto \sum_{j=1}^\ell q_j\sum_{x_k \in S_A} \rho_A(x_k)f(x-x_k\,|\,b_j).
\end{align*}

We evaluate the inner sums for all $x \in S_A$ simultaneously in time $\tilde{\cO}(n)$ using \cref{algorithm:T} and \cref{algorithm:S} described below. Doing the remaining arithmetic in the usual way, we evaluate the right hand sides for all $x \in S_A$ in time $\tilde{\cO}(\ell n)$. The final normalization can be done in time $\cO(n)$.

The rest of this section explains \cref{algorithm:T} and \cref{algorithm:S}. Fix $\rho:\R\to\R$ non-negative, supported on $x_1, \dots, x_n$, and such that its values sum to $1$. Define $Q:\R\to\R$ by
\begin{align*}
  Q(y) \coloneqq \sum_{k=1}^n\rho(x_k)f(y-x_k\,|\,b).
\end{align*}

\cref{algorithm:S} takes as input an arbitrary finite set of real numbers $y_1, \dots, y_m$ and computes, in time $\tilde{\cO}(m+n)$, all of the $m$ quantities $Q(y_1), \dots, Q(y_m)$. Define
\begin{align*}
  &Q_L(y) \coloneqq \sum_{x_k \leq y}\rho(x_k)f(y-x_k\,|\,b)\quad\text{and}\quad
  Q_R(y) \coloneqq \sum_{x_k > y}\rho(x_k)f(y-x_k\,|\,b).
\end{align*}

The following observation, which is immediate from the definitions, is the main idea underlying \cref{algorithm:S} and \cref{algorithm:T}.
\begin{observation}
  If $y$ and $\Delta$ are such that $\{x_1, \dots, x_n\} \cap [y, y+\Delta] = \emptyset$, then
  \begin{align*}
    &Q_L(y+\Delta) = e^{-\frac{\Delta}{b}}Q_L(y)\quad\text{and}\quad
    Q_R(y+\Delta) = e^{\frac{\Delta}{b}}Q_R(y).
  \end{align*}
\end{observation}
\begin{algorithm}\label{algorithm:S} Compute $Q(y_i)$ for all $y_i \in \{y_1,\dots,y_m\}$.
\begin{algorithmic}
\State $U \gets \{x_1,\dots,x_n\} \cup \{y_1,\dots,y_m\}$
\State Sort $U$ from smallest to largest.
\State $L \gets 0$
\State $z_0 \gets U[0]$
\For{$z \in U$}
\State $\Delta \gets z - z_0$
\State $L \gets e^{-\frac{\Delta}{b}}L + \tfrac{1}{2b}\rho(z)$
\State $Q(z) \gets L$
\State $z_0 \gets z$
\EndFor
\State Sort $U$ from largest to smallest.
\State $R \gets 0$
\For{$z \in U$}
\State $\Delta \gets z_0 - z$
\State $R \gets e^{-\frac{\Delta}{b}}R$
\State $Q(z) \gets Q(z) + R$
\State $R \gets R + \tfrac{1}{2b}\rho(z)$
\State $z_0 \gets z$
\EndFor
\end{algorithmic}
\end{algorithm}

It is possible to compute the contribution from $Q_R(y_i)$ during the first iteration over $U$. However, doing so requires that the arithmetic be done using $\gg \tfrac{1}{b}(\text{max}\,U - \text{min}\,U)$ bits of precision because $Q_R(y+\Delta)$ grows exponentially in $\Delta$. In almost all applications it'll be the case that $\tfrac{1}{b}(\text{max}\,U - \text{min}\,U) \gg m+n$, and the algorithm won't run in time $\tilde{\cO}(m+n)$.

Define $T:\R\to\R$ by
\begin{align*}
  T(y) \coloneqq \sum_{k=1}^n\rho(x_k)F(y-x_k\,|\,b).
\end{align*}
$T(y)$ can be decomposed into the three sums
\begin{align*}
  T(y) = \sum_{x_k \leq y}\rho(x_k) - \sum_{x_k \leq y}\tfrac{1}{2}\rho(x_k)\exp\!\left(\frac{x_k - y}{b}\right) + \sum_{x_k > y}\tfrac{1}{2}\rho(x_k)\exp\!\left(\frac{y - x_k}{b}\right).
\end{align*}

With this decomposition and the ideas used to produce \cref{algorithm:L_H} and \cref{algorithm:S}, we can construct \cref{algorithm:T} that takes as input a set $\{y_1,\dots,y_m\}$ and computes all of the corresponding values $T(y_i)$ in time $\tilde{\cO}(m+n)$.
\begin{algorithm}\label{algorithm:T} Compute $T(y_i)$ for all $y_i \in \{y_1,\dots,y_m\}$.
\begin{algorithmic}
\State $U \gets \{x_1,\dots,x_n\} \cup \{y_1,\dots,y_m\}$
\State Sort $U$ from smallest to largest.
\State $M \gets 0$
\State $L \gets 0$
\State $z_0 \gets U[0]$
\For{$z \in U$}
\State $\Delta \gets z - z_0$
\State $M \gets M + \rho(z)$
\State $L \gets e^{-\frac{\Delta}{b}}L + \tfrac{1}{2}\rho(z)$
\State $T(z) \gets M - L$
\State $z_0 \gets z$
\EndFor
\State Sort $U$ from largest to smallest.
\State $R \gets 0$
\For{$z \in U$}
\State $\Delta \gets z_0 - z$
\State $R \gets e^{-\frac{\Delta}{b}}R$
\State $T(z) \gets T(z) + R$
\State $R \gets R + \tfrac{1}{2}\rho(z)$
\State $z_0 \gets z$
\EndFor
\end{algorithmic}
\end{algorithm}

\section{Performance in Duelyst I\hspace{-.7pt}I}\label{section:performance}

In this section, we compare the performance of Glicko2 with our system, as well as our system but with $\beta = 0.9$ instead of $\beta = 0.8$ in \eqref{eq:d2_lambda}, on the dataset of the first $1,\!126,\!592$ ranked matches played since Duelyst \rtwo's launch. Duelyst \rtwo\hspace{0cm} used Glicko2 \cite{glicko2} to rate players previously, with parameters chosen to be the same as the ones used in the prequel Duelyst between 2016 to 2020: $\tau = 0.5$ and default rating $1500$, RD $200$, and volatility $0.06$ \cite{duelystglicko}.

Each of the three systems we analyze in this section we processed the matches in our dataset in chronological order. For each match, each system estimated the probability $p$ of the observed match outcome occurring. Our system estimated $p$ using \cref{model:match}, and Glicko2 estimated $p$ using \cite[Eq.\ (16)]{glicko}. For the matches in which both players had variance less than $70^2$ after the reparameterization \eqref{eq:d2_parameterization}, we computed $-\log p$, the \textit{log loss} of that match. The average log loss was $0.6625$ for Glicko2, $0.6613$ for $\beta = 0.8$, and $0.6559$ for $\beta = 0.9$.

\begin{figure}[H]
  \includegraphics[width=\textwidth]{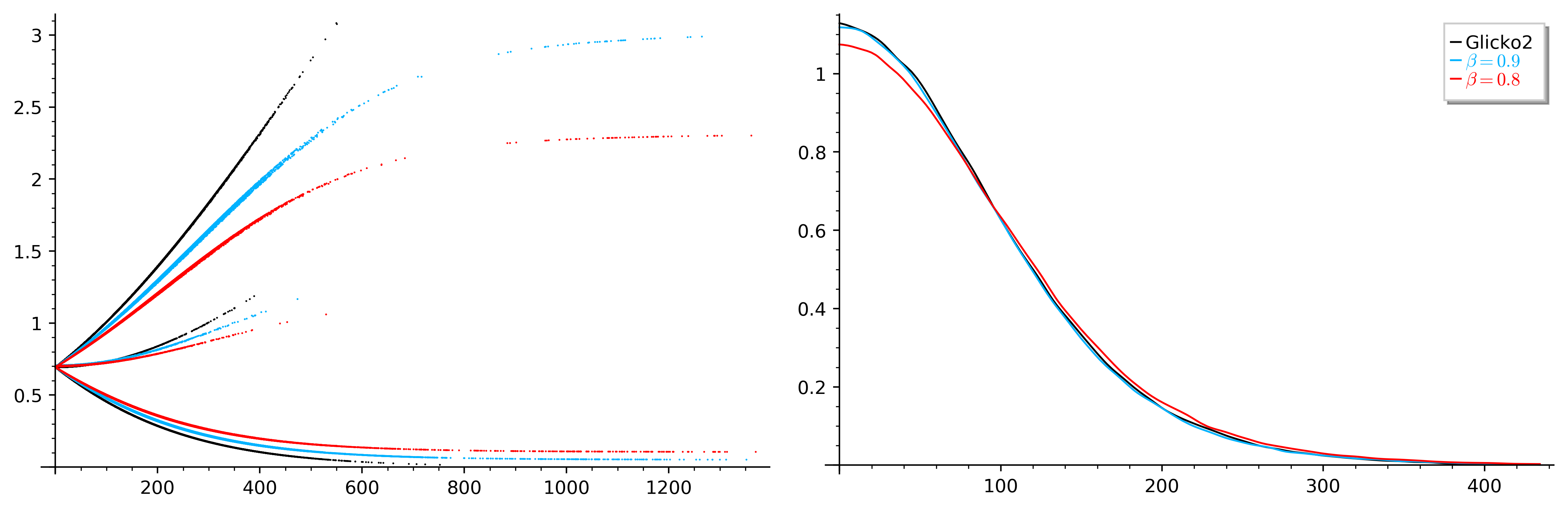}
\end{figure}
\begin{figurecap}\label{fig:logloss}
  Left: Log loss of each match by rating difference. Right: Proportion of the dataset's total log loss by rating difference, normalized to integrate to average log loss.
\end{figurecap}

The left image in \cref{fig:logloss} plots the log loss of each match individually, coloured by system. The horizontal axis is the difference in the ratings of the two players. For each colour, the three distinct curves correspond to wins by the weaker player (top), draws (middle), and wins by the stronger player (bottom). The number of matches being plotted is large: $557973$ for Glicko2, $566213$ for $\beta = 0.8$, and $653660$ for $\beta = 0.9$, causing many of the points to overlap.

The right image of \cref{fig:logloss} quantifies the density of points in the left image by showing the relative contribution of each rating difference to the total log loss in the dataset. Let $$K(x,y) \coloneqq \frac{1}{\sqrt{2\pi\cdot 5^2}} \exp\!\left(-\frac{(x-y)^2}{2\cdot 5^2}\right)$$ denote a Gaussian kernel with variance $5^2$. The curve plotted on the right is proportional to
\begin{align*}
  \sum -\log p \cdot \frac{K(x, |r_A - r_B|)}{\int_0^\infty K(x, t)\, dt},
\end{align*}
where the sum is over matches in which both players have variance at most $70^2$, the quantities $r_A$ and $r_B$ denote the means of the players (i.e.\ their ratings), $p$ is the probability of the observed match outcome occurring as estimated by each system, and $x$ is the variable for the horizontal axis. The proportionality constant is such that the plotted function integrates to the average log loss.

While the value $\beta = 0.9$ had smaller total log loss than the value $\beta = 0.8$ which was implemented, our judgment was that $\beta = 0.8$ was the better choice for our purposes for two reasons.

First, in our application, maximizing the probability observing the empirical data was not our goal. For us, it was much more important to accurately rank the game's top players relative to each other. The choice $\beta = 0.8$ yields more stable and reliable rankings, which is very important in practice.

Second, many players actively enjoy interacting with the in-game rating system; trying to maximize the number the game displays to them becomes one of their primary objectives. From the perspective of game design, harshly penalizing unlucky losses is remarkably frustrating.

\bibliographystyle{plain}
\bibliography{ratingbib}{}

\end{document}